\documentstyle[11pt]{article}
           \newcommand{\beq}{\begin{equation}}
           \newcommand{\eeq}{\end{equation}}
           \newcommand{\beqa}{\begin{eqnarray}}
           \newcommand{\eeqa}{\end{eqnarray}}

\begin{document}

           \title{Phase Structure of Non-Compact QED3 and the Abelian
Higgs Model \footnote{To be published in the proceedings of the 3rd
International Symposium on Quantum Theory and Symmetries, University
          of Cincinnati, September 10-14, 2003.} }

\author{Thomas Appelquist\\
           Department of Physics, Yale University, New Haven, CT 06511
           \\ \\
           L.C.R. Wijewardhana \\
           Department of Physics, University of Cincinnati, Cincinnati,
           OH 45221}
           \date{May 07, 2004}

           \maketitle

           \vspace{-36pt}

           \begin{abstract}

          We review the phase structure of a three-dimensional,
non-compact Abelian gauge theory (QED3) as a function of the number $N$ of
4-component massless fermions. There is a critical $N_{c}$ up to which there
is dynamical fermion mass generation and an associated global symmetry
breaking. We discuss various approaches to the determination of $N_c$, which
lead to  estimates ranging from $N_c =1$ to $N_c =4$. This theory with $N=2$
has been employed as an effective continuum theory for the 2D quantum
antiferromagnet where the observed Neel ordering corresponds to dynamical
fermion mass generation. Thus the value of  $N_c$ is
     of some physical interest. We also consider the
phase structure of the model with a finite gauge boson mass (the
Abelian Higgs model).

           \end{abstract}

           \newpage

           \section{Introduction}

We review the dynamical generation of fermion mass in a three-dimensional,
non-compact Abelian gauge theory (QED3) and its generalization to the case
of a finite gauge boson mass. Attention was first
            drawn to this problem by Pisarski
\cite{Pisarski}. Some years ago, a study concluded that with the gauge
symmetry unbroken, mass
            generation will occur, corresponding to the breaking of a certain
            global
"chiral" symmetry,
            if and only if the number $N$ of four-component
            fermions is no larger than a critical value $N_c$, estimated to be
            $4$ \cite{ANW,Nash}.
            Since then, most studies have agreed that there is
            indeed such an $N_c$, but there has been little agreement
            as to its magnitude.
            Estimates have ranged from that of Refs. 2 and 3
            to as low as
            $N_{c} = 1$~\cite{ACS}. This uncertainty is typical of our
            rudimentary
            understanding of most strongly coupled quantum field theories.

            The value of $N_c$ is
            a potentially important real-world question. It has been suggested
            that QED3 with $ N = 2$ massless fermions can provide a
continuum description of
             a 2D quantum
             antiferromagnet \cite{HTS2,DSS}.  Marston~\cite{HTS3}, Laughlin
             and Zou \cite{Laughlin}, and others have noted that the
            observed Neel ordering corresponds to dynamical fermion mass
              generation
             and its associated chiral symmetry breaking in QED3 with $N=2$.
             More recent discussions of these ideas are provided by Kim
               and Lee~\cite{KimLee}, and Kleinert, Nogueira, and Sudbo
\cite{kleinert}. For another  application of non-compact $QED3$ to condensed
matter systems see Ref. \cite{tes}.
               The suggestion that (non-compact) QED3 provides a continuum
description of the 2D quantum
             antiferromagnet has, however, been challenged in a recent series of
             papers by Herbut et.al \cite{herbut}.

              Several approaches have been brought to bear on the
             determination of
             $N_c$ in QED3. The original study \cite{ANW,Nash} was based on
a questionable use of the $1/N$ expansion along with a
             continuum gap equation. Continuing to non-integral values
of N, it led
             to the estimate $N_c \approx 128/ 3 \pi^2 \approx 4.3$.
This approach
             also led to some speculation on the nature of the phase
transition as a
             function of $N$ \cite{ATW}, namely that because of the
long-range force
             it is of infinite order. A lower value of $N_c$ is suggested by a
             conjectured inequality
             based on the counting of thermodynamic degrees of
freedom~\cite{ACS}.
             It indicates that $N_c$
             can be no greater than $3/2$. It is likely that the determination
of $N_c$ will
             be settled only by numerical lattice studies. These studies
have so far
             suggested that  $N_c$ could be as low as
             2 \cite{latt1,latt2}, but a conclusive
             determination will need more refined simulations.

       We first review the features of QED3 and discuss its
            symmetry properties and behavior for large $N$ where the
$1/N$ expansion may be reliably employed. We  discuss
      fermion mass generation, which takes place at low values of N. The
conjectured constraint of Ref. 4 is then described and applied to QED3. We
review recent lattice computations, discuss their sensitivity to finite size
effects, and to provide some perspective on these simulations include a
rough numerical estimate using the continuum gap equation. We also discuss
dynamical fermion mass generation in a related theory, the Abelian Higgs
model in three-dimensions with the same fermion content as QED3, but where
the gauge field is massive. We then summarize, revisit the relevance of
these models to condensed matter systems, and describe some open questions.

\section{QED3 and its Symmetry Breaking}

       The Lagrangian of the model is

\begin{equation}
L = \sum_{j=1}^{N} \bar{\psi}_{j}(i\not\!\!D)\psi_{j} - {1 \over 4}
F_{\mu\nu}F^{\mu\nu}, \label{L}
\end{equation}
             where $D_{\mu} = i \partial_{\mu} + e
             A_{\mu} $ is the covariant derivative, $e$ is the gauge
coupling, and $\psi_{j}$ is a set of $N$
             4-component fermion fields. To explore
              the phase structure as a function of $N$, it is
              convenient to keep fixed the dimensionful quantity $\alpha
               \equiv { e^2 N\over 8}$. Being super-renormalizable, this
theory is UV-complete, rapidly damped at momentum scales beyond $\alpha$. As
an effective theory at lower momenta, one could nevertheless take the view
that higher momentum physics is unknown, being integrated out into a tower
of higher-dimension operators associated with the scale $\alpha$. We do not
include such terms , which could modify 
 the critical
value of N for chiral symmetry breaking. We return to this question in the
summary when we discuss possible condensed matter applications of QED3.

For large $N$, the theory remains weakly coupled at all momentum scales.
This can be seen by computing the gauge boson propagator in the large-$N$
limit and extracting from it an effective, dimensionless running coupling
${\bar\alpha(k)}$. The large-N form of the Euclidean propagator is $( k^2 +
\alpha  k)^{-1}$, where $k$ is the magnitude of the Euclidean three-momentum
and the second term arises from the $N$ fermion loops.   Multiplying the
propagator by the coupling $e^2$ and by one power of $k$ to make a
dimensionless quantity, we have
             \begin{equation}
             {\bar\alpha(k)} \equiv \frac{e^{2}k}{k^2 + (e^{2}N/8)k}
             ~~=~~\frac{8}{N}\frac{\alpha}{\alpha +k}. \label{runningcc}
              \end{equation}
This expression exhibits asymptotic freedom at large momentum, and conformal
symmetry with an IR fixed point of strength $8/N$ as $k \rightarrow
0$. For large N, the coupling is always weak, and no dynamical fermion mass
generation is expected.

For finite N, on the other hand, the infrared coupling becomes strong,
governed by a strong IR fixed point, and fermion mass generation becomes
possible.  In the absence of fermion mass, the global symmetry associated
with the N four-component fermions is $U(2N)$. The dynamical generation of
an equal mass for all
            the fermion flavors would break the global symmetry to $U(N)\otimes
            U(N)$ giving rise to $2N^2$ Goldstone bosons \cite{Pisarski}. More
            generally, mass terms for the Dirac spinors can violate
parity (P) and time
             reversal symmetry (T), and the gauge field admits a corresponding
             P- and T-violating
            Chern-Simons mass term. These are not generated
            spontaneously \cite{PT paper}. We concentrate here on the
            parity conserving case.

Several studies over the years have led to the conclusion that a parity
conserving mass is indeed generated dynamically for this theory providing
that $N$ is no larger than a critical value $N_c$ \cite{ANW}. An upper limit
is expected, since, as we have noted, the theory is weakly coupled at all
momentum scales in the large-N limit. Since the Lagrangian of Eq. 1 is
UV-complete, damped rapidly at momentum scales above $\alpha$, the value of
$N_c$ is completely determined by the conformal, IR behavior.

\section{A Conjectured Constraint}

A conjectured constraint on the infrared structure of
asymptotically free gauge theories~\cite{ACS} can be utilized to analyze the
phase structure of QED3. This constraint, which takes the form of an
inequality, states that for a wide class of such theories, the number of
IR "degrees of freedom", defined using the thermal free energy per
unit volume $ F(T)$ of the theory, is less than or equal to the number of
corresponding UV degrees of freedom.

These degree-of-freedom counts are defined by the quantities $ f_{IR}$ and $
f_{UV} $, given in terms of $ F(T)$ by taking the the zero- and infinite-T
limits respectively of the quantity $ - {F(T)\over T^d} {f(d)}$. Here $f(d)$
is a function of the number of space time dimensions d, defined such that
the contribution from a free bosonic field is $1$. The inequality is then
\begin{equation}\label{inequality}
           f_{IR} \leq f_{UV}.
\end{equation}
In general, $ f_{IR}$ and $ f_{UV} $ will exist providing that the theory is
governed by IR and UV fixed points. The conjectured inequality was
restricted, however, to asymptotically free theories, in which $f_{UV}$ can
be computed from free field theory. It has been applied to a
variety of four-dimensional theories, and found to be satisfied whenever
$f_{IR}$ can be reliably computed \cite{ACS}. This includes both
IR-free theories and theories with weak IR fixed points.

For QED3, the asymptotic freedom leads to $f_{UV} = {3\over4}(4N) + 1 $
\cite{ACS}. The second term counts the single bosonic degree of freedom
associated with a massless gauge boson in three space-time dimensions. In
the first term, the $4N$ counts the number of fermionic degrees of freedom
associated with the $N$ four-component fermions, and the $3/4$ represents
the Boltzman weighting of fermions in three space-time dimensions.

The value of $f_{IR}$ depends on whether the global symmetry is
spontaneously broken. For large $N$, breaking is not expected, and only the
massless fermions along with the one gauge degree of freedom remain in the
infrared spectrum. From Eq. 2, one sees that the infrared theory is
described by a weak (O(1/N)) fixed point. The free-field result, $f_{IR} =
{3\over4}( 4N) +1  (= f_{UV})$, can then be corrected perturbatively in
$1/N$, the next-to-leading term being $O(1)$. The computation, similar in
some ways to the corresponding perturbative computation in a 4D gauge
theory, has not yet been done. The 4D result is negative, and if the same is
true of the 3D $1/N$ computation, the inequality $f_{IR} \leq f_{UV}$ will
be satisfied.

Now consider the more interesting possibility that for some finite $N$ the
global symmetry is broken. The fermions become massive and $2N^2$ Goldstone
bosons are formed. The infrared (massless) theory consists   of only the
Goldstone bosons and the gauge boson. Thus $f_{IR} = 2N^2 + 1$ and the
inequality will be satisfied only if $N_c \leq 3/2$. If this upper limit on
$N_c$ is correct, then QED3 cannot be in the broken phase if $N$ is larger
then $3\over 2$, meaning that the continuum gap equation \cite{ANW}
overestimates $N_c$.
%If this is the case, QED3 with $N=2$ cannot
%describe the 2D
%quantum antiferromagnet, since the lack of chiral symmetry breaking would
%mean that the observed Neel ordering could not take place.

A natural question about this application of the inequality $f_{IR} \leq
f_{UV}$ has to do with the Mermin-Wagner-Coleman theorem~\cite{Mermin}. We
have determined $f_{IR}$ in the broken phase by computing the free energy at
a temperature $0 < T << \alpha$, counting the Nambu-Goldstone bosons plus the
gauge field as the relevant, non-interacting degrees of freedom, multiplying
by $1/T^3$ and an appropriate constant, and taking the limit $T \rightarrow
0$. Now at finite $T$, the IR behavior of this theory is that of
the corresponding 2D theory. But the Mermin-Wagner-Coleman theorem states
that there can be no spontaneous symmetry breaking, with its Nambu-Goldstone
bosons, in 2D. This question was addressed by Rosenstein, Warr, and Park
\cite{rwp}, who concluded that the symmetry is indeed unbroken at any
non-zero T, and that for small T the zero-temperature Nambu-Goldstone bosons
develop small masses. In our notation, one finds $m_{NG}^2 \sim T^{2} exp(-
\alpha / T)$, arising from the derivative interactions among the (pseudo)
Nambu-Goldstone bosons. Since this mass vanishes more rapidly than $T$ as $T
\rightarrow 0$, it has no effect on the computation of $f_{IR}$.

\section{Lattice Studies}

Lattice studies may be the most reliable method to analyze dynamical mass
generation and directly determine $N_c$ in QED3. Numerical simulations began
over a decade ago, with a study in the quenched approximation giving
preliminary evidence that the global chiral symmetry is spontaneously
broken~\cite{HandsKogut}. Recent advances in computing power have led to
improved studies of mass generation in QED3. Simulations of 2-flavor QED3 by
Hands, Kogut and Strouthos ~\cite{latt1}  on lattices of up to ${50}^3$
sites report that the condensate is two orders of magnitude smaller than the
quenched condensate value. They find that the value of the dimensionless
condensate, defined by
\begin{equation}
\sigma = {<\bar \psi \psi> \over e^4}, \label{condensate}
\end{equation}
where $e$ is the dimensionful gauge coupling constant, is bounded above by $
5 \times 10^{-5}$. They also analyze finite-size effects and find that this
bound is stable for lattice sizes ranging from $10^3$ to $50^3$. They
conclude that there is no decisive signal for chiral symmetry breaking for
$N \geq 2$.

This conclusion is consistent with the conjectured inequality constraint
described in the previous section, which led to $N_c \leq 3/2$ for QED3. On
the other hand, this interpretation of the lattice results is likely
premature. It has been noted by Gusynin and Reenders \cite{Gus} that even a
lattice of size ${50}^3$ provides a sufficient IR cutoff to affect
substantially the value of $N_c$. They use the continuum gap equation first
employed to note the existence of an $N_c$, modeling the effect of a finite
lattice size by imposing an IR cutoff $\mu$ on the integral in the gap
equation. Because of the scale-invariant form of the gap equation following
from the dominance by the IR fixed point, the equation is logarithmic in
character. A measure of the importance of the cutoff $\mu$ relative to the
effective UV cutoff $\alpha$ is therefore $ln ~\alpha / \mu$. As this
quantity decreases from infinity, a growing portion of momentum space is
truncated, and therefore a stronger gauge coupling is required to break the
symmetry. Thus, $N_c$ should drop. Gusynin and Reenders find, for example,
that with $ln~\alpha / \mu$ still as large as $6$ ($\alpha / \mu \approx
400$), $N_c$ drops by more than $30\%$. A lattice of size ${50}^3$
corresponds to a much smaller $\alpha / \mu$, indicating that chiral
symmetry breaking should not take place for $N=2$.

If this analysis is qualitatively correct, that is, if the continuum IR
cutoff correctly models the finite size of a lattice, then only
a larger lattice would be able to determine accurately the value of $N_c$ for a
theory such as QED3, governed by a (conformal) IR fixed point. This same
remark would apply to studies of the conformal phase transition as a
function of the number of fermion species in four-dimensional gauge
theories.

\section{Gap Equation Estimates}

Suppose that it becomes possible to carry out simulations with such large
lattices that finite size effects are no longer important, and suppose that
the results of the simulations for $\sigma$ (Eq. 4) with $N=2$ continue to
be bounded as in Ref. \cite{latt1}. What would one conclude? Here we present
a rough argument using the continuum gap equation indicating that even then
it could be unclear whether $N=2$ is in the symmetric or broken phase.

The spirit of the gap equation approach is to use the large-N form of the
kernel of this equation (the large-N gauge-boson propagator ), and then to
continue to finite $N$. The reliability of this approach, which first
suggested the existence of a finite $N_c ~ (\simeq 128/3\pi^2)$, is not
clear because the higher terms in the kernel are not parametrically small.
There is, however, some evidence that the corrections are small
numerically~\cite{Nash}. We use this approach to estimate the condensate for
$N = 2$, expecting only that the estimate is order-of-magnitude.

The theory is UV-complete, rapidly damped at momentum scales beyond
$\alpha$. Thus, to a good approximation, the gap equation can be written
down employing the $\alpha \rightarrow \infty$ form of the large-N kernel,
with $\alpha$ then used as a UV cutoff. The resulting equation is
\begin{equation}
\Sigma(p) = \frac{16}{3N \pi^{2} p} \int^{\alpha} \frac{k dk \Sigma(k)}{k^2
+ \Sigma^{2}(k)}
            \left[ p+k -|p-k|\right],
\label{Sigma}
\end{equation}
where we make use of the non-local Kondo-Nakatani gauge for which the
leading large-N form of the wave function renormalization is unity. The
critical N determined from this equation is $N_c = {128 / 3\pi^2}$.

Numerical and analytical solutions  of this equation ~\cite{ABCW}  show that for a range of N such that $ {{ N_c} \over { N } } >
1$,  the overall scale of the solution $\Sigma(p) $ of Eq.(5),  set by $\Sigma(0) $,  is much less than
$ \alpha $,     and 
for $ N=2 $,  
\begin{equation}
  {\Sigma(0) \over \alpha} \approx 1.7\times 10^{-2}. 
\end{equation}
    In the range of momentum $ \Sigma(p) < p < \alpha $ the solution
has the following approximate form 
\begin{equation}
\Sigma(p)  = {{\Sigma(0)}^{3\over2}\over  p^{ {1\over 2} } }~sin [{{1\over
2}\sqrt {( {N_c\over N} -1 )} (ln({{ p\over\Sigma(0)}) +\delta})}],
\label{Sol}
\end{equation}
where $\Sigma(0)$ is used to scale the log and $\delta$ is a phase.

We use these formulae to estimate the dimensionless fermion condensate of
Eq. \ref{condensate}. The fermion condensate $<\bar \psi \psi >$ is given by
the integral
\begin{equation} <\bar \psi \psi> =  \int^{\alpha} {d^3 k\over
({2\pi})^3 } { 4\Sigma(k) \over{ k^2  } } \\ =  {2\over  \pi^2} \int^{\alpha} dk \Sigma(k).
\label{dimfulcondensate}
\end{equation}
Noting that the sine function is bounded above by $1$ in the region of
integration up to $\alpha$ , we see that the dimensionless condensate
$\sigma$ of Eq. \ref{condensate} ($ = <N ^{2}\bar \psi \psi /
         64 \alpha^2> $) is bounded above by
$$ {N^2 \over 16 \pi^2} ({\Sigma(0) \over
\alpha})^{3\over2}.  $$  With $ N = 2 $,  $N_c = {128 / 3\pi^2}$ and 
 $ {\Sigma(0) \over
\alpha} \approx 1.7\times 10^{-2} $,  this expression becomes $ 6 \times 10^{-5} $ which is slightly above the 
upper bound of $ 5 \times 10^{-5}$ found in lattice simulations ~\cite{latt1}.  
We have numerically estimated the actual value of the 
condensate to be 
 $ 4.4 \times 10^{-5} $ which falls  below  the lattice bound~\cite{Valery}.

This suggests that a dimensionless condensate (\ref{condensate}) of order
$10^{-5}$   may naturally arise in QED3 for N of order 2 even when
$N_c$ is as large as $128/3\pi^2 \approx 4$. It indicates that even if the
numerical simulations are performed on a very large lattice, so that
finite size effects are unimportant, a   precise computation will be
required to decide whether $N = 2 $ is in the broken or symmetric phase.

\section{Abelian Higgs Model}

       We conclude by discussing fermion mass generation when the
Abelian gauge field has a non-zero mass. This theory has been used as a
continuum description of the competition between long range
antiferromagnetic order and superconducting order in planar cuprate
systems~\cite{liucheng}, \cite{PBF}. It can also be used to interpolate
between QED3 and the 2+1 dimensional Thirring model, the latter having been
studied in recent lattice simulations ~\cite{latt1}. ( Related studies,
using continuum gap equation methods and invoking the idea of hidden local
gauge symmetry may be found in references \cite{Koreans} \cite{Sugiyura} ).
Finally, it can be used to study the effect of an IR cutoff on QED3
\cite{PBF}, similarly to the use of a simple IR cutoff as described in
Section 3 \cite{Gus}.

        The Lagrangian of the   model  is
            \begin{equation}
            {\cal L} = {\cal L}_{QED3} + {1\over 2} D_{\mu}\Phi^{*} D^{\mu}
              \Phi
             - \lambda(\Phi^{*} \Phi-N v)^{2}
             \label{L}
             \end{equation}
            where ${\cal L}_{QED3}$ is  given in equation (1),
       and $\Phi$ is a complex
scalar field
            with vacuum expectation value $v$.  The  addition of
the scalar field does not change the global symmetry structure of the
theory. We remove the Higgs boson from the spectrum by taking the limit
             $\lambda \rightarrow \infty$ with $v$ fixed. This leaves
the theory UV-complete. The gauge boson mass is $M = e\sqrt{N v}$. To
explore
             the phase structure  as a function of $N$, we
             keep fixed the quantities $\alpha
              \equiv { e^2 N\over 8}$ and $v^{2}$. The gauge boson
             mass $M$ is then also fixed.

            As in QED3, we define a dimensionless running gauge
coupling ${\bar\alpha(k)}$. The coefficient of  $g_{\mu \nu}$ in the
Euclidean gauge-boson propagator takes the form $( k^2 + M^2 + \alpha
k)^{-1}$ to leading order in $1/N$, where $k$ is the magnitude of the
Euclidean three-momentum and the last term arises from the $N$ fermion
loops. The $k_{\mu} k_{\nu} $ term of the propagator is dropped since it
couples to a conserved current. Multiplying the propagator by the coupling
$e^2$ and by   $k$ to make it dimensionless  , we have
            \beq
            {\bar\alpha(k)} \equiv \frac{e^{2}k}{k^2 + M^2 + (e^{2}N/8)k}
            ~~=~~\frac{8}{N}\frac{k}{(k^{2}/\alpha)+ G^{-1} +k},
\label{runningcc}
             \eeq
where $G \equiv \alpha/M^{2}$. This effective coupling vanishes in
both the UV and IR
limits. It reaches a maximum no larger than $8/N$, attainable only when $M
<< \alpha$.   As
$M\over \alpha $ increases, the maximum value decreases
monotonically. Thus, for
any $M/\alpha$, the model is
weakly coupled in the large-N limit at all momentum scales.

It is reasonable to expect that, as in QED3, a parity conserving mass is
      generated  for $N$ below some critical value
$N_c$. Assuming an $N_c$ to exist, what can it depend on? Since the forces are
strongly damped at scales beyond $\alpha$, no  additional cutoff is
necessary to define the theory. Thus $N_c$ can be a function of only the
dimensionless ratio $M/\alpha$. This  dependence describes the
boundary of a phase diagram for the theory.

Consider first the limit $M/\alpha \rightarrow 0$. On the one hand, this may
be thought of as taking $M \rightarrow 0$ with $\alpha$ fixed, giving QED3.
As discussed earlier, estimates of $N_c$ in this limit have ranged from
$3/2$  to approximately  $4$.  The limit $M/\alpha \rightarrow 0 $ may also
be taken by letting $\alpha \rightarrow \infty$ with $G \equiv \alpha/M^2$
fixed. This leads to the 2+1 dimensional Thirring model, meaning that $N_c $
for this model is the same as that for QED3. This assumes that the Thirring
model is treated with any UV cutoff taken very large relative to $G^{-1}$.
For finite $N$, this can be implemented using lattice techniques. (For large
$N$, these two ways of taking the limit $M/\alpha \rightarrow 0$ may be
considered by examining ${\bar\alpha}(k)$ (Eq. 11). In the case
$M\rightarrow 0$ with $\alpha$ fixed (QED3),  there appears an IR ($k <<
\alpha $) fixed point of strength $8/N$. In the case $\alpha \rightarrow
\infty$ with $G^{-1}$ fixed , there is a UV ($k >> G^{-1}$) fixed point of
the same strength).

As $M/\alpha$ is increased from zero, low momentum components are damped
out. If the force driving the symmetry breaking is attractive at all scales
(as in the large N approximation),  a stronger infrared coupling is required
to trigger dynamical symmetry breaking, and therefore $N_c$ will decrease. At
some value of $M/\alpha$, $N_c$ drops below unity meaning that symmetry
breaking will not take place at all ~\cite{liucheng}. This critical curve
should be similar to that determined by Gusynin and Reenders \cite{Gus},
with the gauge boson mass replacing the explicit IR cutoff. The character of
the phase transition along this critical curve is also of interest. For
$M/\alpha = 0$, as discussed earlier, because the force is of infinite
range, the continuum gap equation suggests that the transition is of
infinite order \cite{ATW}. With $M > 0$, the force becomes of finite range,
and the transition is of second order.

Does the conjectured inequality $ f_{IR} \leq f_{UV} $ provide any
information about $N_c$ as a function of $M/\alpha$? With $\alpha \equiv {
e^2 N\over 8}$ finite, the theory is asymptotically free. A free-field
computation then gives $f_{UV} = {3\over4}( 4N) + 2 = 3N +2 $~\cite{ACS}.
The second term counts the $2$ bosonic degrees of freedom associated with a
massive gauge boson in three space-time dimensions. As in QED3,  the value
of $f_{IR}$ depends on whether the global symmetry is spontaneously broken.
For large $N$, breaking is not expected,  and, if $M$ is nonzero, the
infrared theory consists of only the (non-interacting) massless fermions.
Then $f_{IR} = {3\over4}( 4N)$, and the inequality is satisfied. Now suppose
that for some finite $N$ the global symmetry is broken. The fermions become
massive, and for finite gauge boson mass $M$ the IR theory consists of only
the $2N^2$ non-interacting Goldstone degrees of freedom. Thus $f_{IR} =
2N^2$,  and the inequality demands that $N_c$ be no greater than $2$. But
the inequality already demands that $N_c$ be no greater than $3/2$ when $M =
0$, and we expect that $N_c$ will decrease as $M / \alpha$ increases. So it
would seem that the conjectured inequality has nothing useful to say about
the shape of this critical curve.

Lattice simulations of the Abelian Higgs model for an appropriate range of
values of $M/\alpha $ could determine the shape of this critical curve.
Simulations have  been done only for QED3 and the Thirring model, which
correspond to the limit $M/\alpha \rightarrow 0$. We discussed in Section 3
the sensitivity of the QED3 simulation to finite size effects. A  
numerical study of the Thrring model was reported by Hands and
Lucini~\cite{hands}. We noted above that providing this model is treated
with a  UV cutoff large compared to the inverse four-fermion
coupling $G^{-1}$, the value of the critical coupling $N_c$ should approach
that of QED3. The lattice simulations of Ref. \cite{hands} do   take
the ultraviolet cutoff larger than  $G^{-1}$,  and find chiral symmetry
breaking   for $N\approx 3$.
Their results remain sensitive to the UV cutoff, but we anticipate
that $N_c$ will only increase as the cutoff is made larger (as the continuum
limit is approached).  This suggests that $N_c > 3 $ for the continuum
Thirring  model, and therefore also for QED3.

We end this section by digressing
      to use the large-N Thirring model to
explain why the  inequality~\cite{ACS} was  restricted to asymptotically
free theories. The model becomes non-interacting in the infrared ($k <<
G^{-1}$) so $f_{IR} = 3N$. Having taken $\alpha \rightarrow \infty$, the
resultant weak UV fixed point means that $f_{UV}$ exists and may be computed
in the $1/N$ expansion. To leading order, $f_{UV} = 3N$. The computation of
the next-to-leading corrections should be the same as that of $f_{IR}$ in
QED3, since the latter is governed by an IR fixed point identical in
strength to the UV fixed point of the Thirring model. If the result is
negative as anticipated, then $f_{IR} > f_{UV}$ for this model. Another 2+1
dimensional model with a UV fixed point, leading to $f_{IR} > f_{UV}$, was
studied by Sachdev \cite{sachdev2}. In light of
      these  examples, the conjectured inequalty was restricted to
asymptotically free theories.

\section{Summary and Discussion}

        A key question for a non-compact Abelian gauge theory in three
space-time
        dimensions is the critical number $N_{c}$ of
four-component fermions below which there is fermion mass generation with an
associated chiral symmetry breaking, and above which the fermions remain
massless. Over the years, estimates of $N_{c}$ with the gauge symmetry
unbroken (QED3) have ranged from infinity to one. Recent studies  agree that
it is finite and about $4$, but some estimates continue to be as
low as $1$. Since this theory is UV-complete, damping rapidly at momentum
scales above $\alpha \equiv e^{2} N / 8$, the value of $N_c$ depends only of
the IR behavior of the theory, where conformal symmetry sets in.

This question has a potential real-world interest, because of the
possibility that QED3 with $N = 2 $ can describe the physics of the
      planar anti-ferromagnet ~\cite{HTS2} ~\cite{DSS}. The observed
antiferromagnetic (Neel) ordering
      corresponds to chiral symmetry breaking in QED3 ~\cite{HTS3}
~\cite{Laughlin}. Numerical simulations~\cite{AF} of the planar
antiferromagnet also indicate Neel ordering.
%Thus, if the QED3 description of this system is to be valid,
%it must be the case that $N_c > 2$ for QED3.
     Herbut and collaborators \cite{herbut} have
challenged the idea that non-compact QED3 is relevant to the planar
antiferromagnet, noting that the underlying  theory is compact QED3 with its
topological (instanton) structure. Some argue, however, that for large
enough N,  instantons and anti-instantons are bound at distance scales of
order $1/\alpha$ by a logarithmic term in the action~\cite{kleinert}. Then
for $N = 2$ the non-compact theory could be employed at
larger distances. But Herbut et al~\cite{herbut} claim that the finite
density of instantons and anti-instantons screen the logarithmic term in the
action and that instantons are therefore unsuppressed at all N. If this is
the case, the non-compact theory cannot be used for this condensed matter
system~\cite{sachdev}.

In general, the question of whether the value of $N_c$ as determined by
(UV-complete) QED3 is relevant to condensed matter systems depends on the
role of new, short-distance physics. With the physical lattice spacing of
order $1/\alpha = N / 8 e^{2} $, physics at this scale can be represented in
the effective theory at momentum scales below $\alpha$ by the addition of a
set of higher-dimension operators. The importance of these terms depends on
their strength. We have analyzed this question in the framework of the
continuum gap equation in 4-dimensions \cite{ATW2}, which has much the same
structure as QED3 treated in the $1/N$ expansion. With a four-fermion
additive term, there is a critical four-fermion coupling (attractive) of
order unity above which this term by itself can trigger spontaneous chiral
symmetry breaking. As long as the magnitude of this coupling, either
attractive or repulsive, is less than one quarter of its critical value
~\cite{ASTW, carena}, the additive term is unimportant, not affecting the
value of $N_c$. This limit may not be quantitatively accurate, but
it seems reasonable that a limit of this order exists. Whether this limit is satisfied for condensed
matter systems such as the planar antiferromagnet is not clear.  Thus even if the non-compact
theory does provide a continuum descriptiion of the antiferromagnet, the
value of $N_c$ for this system may or may not  be determined accurately
by QED3 itself.

Returning to our summary of QED3, we have briefly reviewed the various
approaches to the determination of $N_{c}$. A recently conjectured
inequality constraining asymptotically free quantum field
theories~\cite{ACS}, when applied to QED3, suggests that $N_c \leq 3/2$.
This bound disagrees with the  estimate $N_c\approx 4 $
emerging from the continuum gap equation approach which originally suggested
the existence of a finite $N_c$.

  Recent lattice simulations
of $N=2$ QED3~\cite{latt1} indicate that the fermion condensate is very
small, possibly signaling that $QED3$ is in the symmetric phase, and
therefore that $N_c < 2$. This conclusion depends, however, on the importance of
finite-size effects. A recent analysis of Gusynin and Reenders ~\cite{Gus}
using the continuum gap equation indicates that  a lattice size of $ 50^3$,  used in the
simulations, provides a strong enough suppression of IR physics to reduce
$N_c$ below 2, thus accounting for the obseravation of   a small upper
bound for the condensate. Using the solution of the gap equation with no IR
cutoff, we have made a rough estimate of the fermion condensate, assuming
that $N_c$ is larger -- approximately $4$ as suggested by the continuum gap
equation studies ~\cite{Nash} . Interestingly, even though $N=2$ is then
well into the broken phase, the condensate is estimated to be very small,
below the upper bound of Ref. \cite{latt1}. This indicates that even if  
simulations are conducted on a large enough lattice to make finite size
effects unimportant, a  precise determination of the condensate  is
needed to decide whether QED3 with $N=2$ is in the symmetric or broken
phase.

The value of $N_c$ in the Abelian Higgs model is also of interest. This
model leads to either QED3 or the 2+1-dimensional Thirring model in the
limit $M/\alpha \rightarrow 0$ depending on how the limit is taken. But
since $N_c$ depends on only the ratio $M/\alpha$ the value of $N_c$ must be
the same in both models.  
Lattice simulations of the
Thirring model on a
$12^3$ lattice lead  to  $N_c\approx 3$~\cite{hands}. This value remains sensitive to the UV cutoff (the
lattice spacing), and should
increase toward the $N_c$ of QED3  as the continuum limit is approached.
On the other hand the inequality (\ref{inequality})  applied to the Abelian Higgs model
yields $N_c \leq 2$.

         To summarize, after more than fifteen years there is
still no definitive answer to the question of the critical number of
four-component fermions in QED3 (and its generalization to the Abelian Higgs
model) marking the boundary between broken and unbroken chiral symmetry.
This question is of interest because it could be relevant to the behavior of
certain condensed matter systems and because the answer requires an
understanding of a strongly coupled, UV-complete quantum field theory.

                                         \section{Acknowledgements}

The work described here was supported in part by funds provided by the U.S.
Department of Energy under Contracts DE-AC02-76ER0 3075 and
DE-FG02-84ER40153. We thank Simon Hands and Subir Sachdev for valuable
conversations, Frank Pinski for his assistance with computing and V. P. Gusynin for helpful correspondence.


\begin{thebibliography}{99}
\bibitem{Valery}V.Gusynin has also obtained a similar estimate for the two flavor condensate, using analytical solutions to the gap
equation described  in Ref. ~\cite{Gus}.
\end{thebibliography}
                                        \end{document}